\documentclass[aps,prb,twocolumn,showpacs,superscriptaddress,amsmath,amssymb]{revtex4-1}

\usepackage{graphicx}
\usepackage{bm}
\usepackage{color}
\renewcommand{\vec}[1]{\bm{#1}}

\begin{document}


\title{Dynamics and relaxation in spin nematics}



\author{V. G. Bar'yakhtar}
\affiliation{Institute of Magnetism, National Academy of Sciences and
  Ministry of Education,  03142 Kiev, Ukraine}

\author{V. I. Butrim}
\affiliation{Taurida National V.I. Vernadsky University, 95007 Simferopol, Ukraine}

\author{A. K. Kolezhuk}
\affiliation{Institute of High Technologies,  Taras Shevchenko National
  University of Kiev, 03022 Kiev, Ukraine}
\affiliation{Institute of Magnetism, National Academy of Sciences and
  Ministry of Education,  03142 Kiev, Ukraine}

\author{B. A. Ivanov}
\affiliation{Institute of Magnetism, National Academy of Sciences and
  Ministry of Education,  03142 Kiev, Ukraine}
\affiliation{Radiophysics Department, Taras Shevchenko National
  University of Kiev, 03022 Kiev, Ukraine}

\date{\today}

\begin{abstract}
We study dynamics and relaxation of elementary excitations (magnons)
in the spin nematic (quadrupole ordered) phase of $S=1$ magnets.  We
develop a general phenomenological theory of spin dynamics and
relaxation for spin-$1$ systems.  Results of the phenomenological
approach are compared to those obtained by microscopic calculations
for the specific $S=1$ model with isotropic bilinear and biquadratic
exchange interactions. This model exhibits a rich behavior depending
on the ratio of bilinear and biquadratic exchange constants, including
several points with an enhanced symmetry.  It is shown that symmetry
plays an important role in relaxation.  Particularly, at the SU(3)
ferromagnetic point the magnon damping $\Gamma$ depends on its
wavevector $k$ as $\Gamma\propto k^{4}$, while a deviation from the
high-symmetry point changes the behavior of the leading term to
$\Gamma\propto k^{2}$.  We point out a similarity between the behavior
of magnon relaxation in spin nematics to that in an isotropic
ferromagnet.
\end{abstract}



\maketitle

\section{Introduction}
\label{sec:intro}

Exchange interaction between atomic spins in solids at low
temperatures usually leads to magnetic ordering, that entails
spontaneous breaking of the time reversal
symmetry. A large body of results in physics of magnetism have been
obtained in the framework of a
phenomenological theory, which describes the state of a magnet at low
temperature by a constant length magnetization vector $\vec{M}$
proportional to the average value of the spin operator
$\langle\widehat{\vec{S}}\rangle$
(or, in the more general case of a magnet
with $n$ sublattices, by several sublattice magnetization vectors
$\vec{M}_\alpha$, $\alpha =1,\;2,...\;n$).\cite{spinwaves,Turov+book,KIK-books}
This approach naturally leads to the macroscopic description of the
spin dynamics by means of the Landau-Lifshitz equations  for the
magnetization vector (or sublattice magnetizations).

The assumption that the magnitude of the magnetization vector is conserved, $|\vec{M}| =\mathrm{const}$,
is justified  in the case of a well pronounced long-range magnetic
ordering with $|\vec{M}|$ being not much different from its saturation
value. In that case longitudinal oscillations that change
$|\vec{M}|$ have much higher energies than  transversal modes
described by the Landau-Lifshitz equations, and thus the longitudinal
modes can be safely neglected in the low-energy, long-wavelength
limit. However, the longitudinal modes may come down and become
comparable in energy with the transversal ones close to phase
boundaries, where thermal or quantum fluctuations lead to a strong
``spin contraction'' so that
 $|\langle\widehat{\vec{S}}\rangle| \ll S$. In that case there might
be other low-energy degrees of freedom as well, for instance, those
related to the so-called spin nematic
ordering\cite{AndrGrishchuk,Papanicolaou1988,Chubukov1990,Chubukov1991}
that is described by the deGennes tensor order
parameter
\[
Q^{ab}= \frac{S(S+1)}{3}\delta_{ab}-\frac{1}{2}\langle
S^{a} S^{b} +S^{b} S^{a} \rangle
\]
 built from quadrupole averages.
Although the latter expression makes sense only for $S>\frac{1}{2}$,
nematic order is possible in spin-$\frac{1}{2}$ systems as
well;\cite{Shannon+06,vekua07,Hikihara+08,Sudan+09} in that case  the
$S>\frac{1}{2}$ spin operators $S^{a}$ above are understood as
composed from $S=\frac{1}{2}$ spins belonging to different lattice sites.
In
a usual magnetically ordered state, spin nematic (quadrupolar) order
$Q^{ab}$ is trivially
present as a ``slave'' of the primary magnetic (dipolar) order $\vec{M}$. In
contrast to that, in a purely nematic state  the average (sublattice)
magnetization $\vec{M}$ is absent even at zero temperature, and the
only order is characterized by nontrivial quadrupole averages.  The
time reversal symmetry remains unbroken in the spin nematic phase.

The static and dynamic properties of spin nematics  have attracted
much interest of researchers during last two
decades.\cite{LoktevOstrovskii94,Fath1995,MikushMosk02,IvanovKolezhuk03,Buchta+05,Sadler2006,Mukerjee+06,Shannon+06,GroverSenthil07,Harada2007,Hikihara+08,BIKK08,BIK10,Sudan+09,BernatskaHolod09,Toth+10,RAKSV11,DeChiara+11,Kovalevskii11}
This interest has received a new boost recently, mainly in the context
of spinor Bose-Einstein condensates (BECs).\cite{KawaguchiUeda12}
Optically trapped ultracold gases provide a unique highly controllable
environment opening an exciting route for simulating a wide range of
strongly correlated systems including quantum magnets.
Ongoing
experiments~\cite{Trotzky+08,Lin+09,Kim+10,Jordens2010,Ma+11,Simon+11,Struck+11}
have already managed to reach the regime suitable to study magnetic
properties of such systems.
Spinor gases are especially interesting because their internal degrees of freedom result in a rich physics, providing
an opportunity to study quantum magnets with strong
non-Heisenberg interactions.

Spin-$1$ gas represents the simplest bosonic spinor system.
Depending on interparticle interactions\cite{Ho1998,Ohmi1998}
determined by
the $s$-wave scattering lengths $a_{0,2}$ for the collision
channels with the total spin $0$ and $2$,
 spin-$1$ BEC has ferromagnetic ground state for $a_0>a_2$ (as
in $^{87}$Rb, see Ref.\ \onlinecite{Barrett2001}), and nematic (polar) ground
state for $a_2>a_0$ (as in $^{23}$Na, see Ref.\ \onlinecite{Stenger1998}).
For spin-$1$ atoms loaded in an optical lattice a variety of phases
has been predicted. \cite{Demler2002,Yip2003,Zhou2003,Imambekov2003,Imambekov2004,Snoek2004,Rizzi2005,Harada2007,Chung2009}

In the Mott insulator regime (intersite hopping much
smaller than the on-site interaction), at low energies spin-$1$ bosons on a lattice can
be
effectively described by a purely spin model  (other
degrees of freedom are separated by a large energy gap). At odd
fillings, this effective model describes $S=1$ spins.  Exchange
interaction of those spins,\cite{Yip2003,Imambekov2003} 
in addition to
the usual Heisenberg exchange terms $(\vec{S}_{i}\cdot\vec{S_{j}})$,
includes strong biquadratic exchange of the type
$(\vec{S}_{i}\cdot\vec{S_{j}})^{2}$. In  three-dimensional
systems, this biquadratic exchange leads
to a long-range-ordered spin-nematic state  for $a_2>a_0$, while the case $a_0=a_2$
exhibits an enlarged $SU(3)$ symmetry with a highly degenerate ground
state.\cite{Batista2002}

In this paper we develop a general phenomenological theory of spin dynamics
and relaxation in isotropic spin-$1$ magnets, focusing on the
properties of the nematic phase in three dimensions.
We compare the results of
the phenomenological approach to those obtained by microscopic calculations for the
$S=1$ lattice model with isotropic bilinear and biquadratic exchange
interactions.  Our goal is to demonstrate the role of enhanced
symmetry in the relaxation.  Particularly, we show that at the $SU(3)$
ferromagnetic point the magnon damping $\Gamma$ depends on its
wavevector $k$ as $\Gamma\propto k^{4}$, while a deviation from the
high-symmetry point changes this behavior to $\Gamma\propto k^{2}$. Our
formalism reveals parallels between the general equations describing
spin dynamics and relaxation in spin nematics and those
proposed earlier by one of the authors \cite{Baryakhtar84} for the dynamics
and relaxation of magnetization in ferromagnets.

The structure of the paper is as follows: in Sect.\ \ref{sec:phenom}
we describe the phenomenological theory of spin dynamics and
relaxation, based on the Onsager relations, as it has been done
earlier for
ferromagnets.\cite{Baryakhtar84} In Sect. \ref{sec:micro}, we discuss
the microscopic calculation of magnon relaxation for the $S=1$
bilinear-biquadratic model, and compare the results of the
phenomenological and microscopic approaches.  Finally,
Sect.\ \ref{sec:summary} contains a brief summary.

\section{Phenomenological theory\\ of spin dynamics and relaxation\\ in $S=1$ magnets}
\label{sec:phenom}

\subsection{Description of spin-1 states and order parameters}

To describe the dynamics of spin-1 system, it is convenient to use the
formalism of $SU(3)$ coherent states.\cite{IvanovKolezhuk03}
The most general pure spin-1 state
$|\psi\rangle_{j}$ at a single given site $j$ is a linear superposition of
three basis states $|\sigma\rangle_{j}$ with
$S_j^{z}|\sigma\rangle_{j}=\sigma|\sigma\rangle_{j}$,
$\sigma=0,\pm1$. It is convenient to write down this state in a
 ``cartesian'' basis of  states
\begin{equation}
\label{states}
|\psi\rangle_{j}=\sum_{a=1,2,3}
z_{ja}|t_{a}\rangle_{j},
\end{equation}
where
$|t_{1}\rangle=\frac{1}{\sqrt{2}}(|-1\rangle-|+1\rangle)$,
$|t_{2}\rangle=\frac{i}{\sqrt{2}}(|-1\rangle+|+1\rangle)$,
$|t_{3}\rangle=|0\rangle$,
then  the object
$\vec{z}_{j}=(z_{j1},z_{j2},z_{j3})$ transforms as a vector
under usual (SU(2)) rotations.

Instead of the three-component complex vector $\vec{z}$, one can use a
different spin-$1$ state parametrization\cite{Ercolessi+01,IKK08}  through
the eight-component real vector $\vec{n}$, defined as follows:
\begin{equation}
\label{8-vector}
n_{\alpha}=z^{*}_{a}\lambda_{ab}^{\alpha} z_{b},
\end{equation}
where $\lambda^{\alpha}$, $\alpha=1,\ldots 8$ are the Gell-Mann
matrices that are hermitean and have the following properties:
\begin{eqnarray}
\label{Gell-Mann}
&& \lambda^{\alpha} \lambda^{\beta}=\frac{2}{3}\delta_{\alpha\beta}\openone
+(d_{\alpha\beta\gamma}+if_{\alpha\beta\gamma})\lambda^{\gamma},\quad
\nonumber\\
&& \lambda^{\alpha}_{ab}\lambda^{\alpha}_{a'b'}=2\delta_{ab'}\delta_{a'b}-\frac{2}{3}\delta_{ab}\delta_{a'b'}.
\end{eqnarray}
Here the tensor of
structure constants $f_{\alpha\beta\gamma}$  is totally
antisymmetric with respect to the permutation of any pair of indices, while
the other tensor $d_{\alpha\beta\gamma}$
is totally symmetric under such operations, see the Appendix.

The octet $\vec{n}$ transforms according to the adjoint representation
of the SU(3) group under  a general unitary (SU(3)) transformation of
the triplet of basis spin-1 states. For two octets   $\vec{n}$ and
$\vec{n'}$, one can define a scalar product $(\vec{n}\cdot
\vec{n'})$,  vector
crossproduct $(\vec{n}\wedge \vec{n'})$, and symmetric vector product
$(\vec{n} * \vec{n'})$ as follows:\cite{Macfarlane+68}
\begin{eqnarray}
\label{octet-prod}
\vec{n}\cdot
\vec{n'}=n_{\alpha}n'_{\alpha}, \quad (\vec{n} *
\vec{n}')_{\alpha}=d_{\alpha\beta\gamma}n_{\beta}n'_{\gamma},&&\nonumber\\
(\vec{n}\wedge\vec{n}')_{\alpha}=f_{\alpha\beta\gamma}n_{\beta}n'_{\gamma}.&&
\end{eqnarray}
For any octet $\vec{n}$, quantities
\begin{equation}
\label{invariants}
I_{2}(\vec{n})=\vec{n}\cdot\vec{n},\quad I_{3}(\vec{n})=\vec{n}\cdot (\vec{n} * \vec{n})
\end{equation}
remain invariant under SU(3) basis tranformations $\vec{z}\mapsto
U\vec{z}$, and satisfy the constraint
\begin{equation}
\label{inv-constr}
I_{2}(\vec{n})^{3} \geq 3 I_{3}(\vec{n})^{2} .
\end{equation}

The density matrix $\widehat{\rho}$ of a single spin can be expressed
through the octet $\vec{n}$:
\begin{equation}
\label{dmatrix-n}
\langle t_{a}| \rho| t_{b}\rangle =\frac{1}{3}\delta_{ab} +\frac{1}{2} n_{\alpha}\lambda^{\alpha}_{ab}.
\end{equation}
For a normalized pure state, the condition $\widehat{\rho}^{2}=\widehat{\rho}$ translates
into the following constraints on $\vec{n}$:
\begin{equation}
\label{pure-constr}
 \vec{n}^{2}=\frac{4}{3},\quad (\vec{n} * \vec{n})=\frac{2}{3}\vec{n},
\end{equation}
i.e., $I_{2}(\vec{n})=4/3$ and $I_{3}(\vec{n})=8/9$, so
(\ref{inv-constr}) becomes an equality.
One can show that the
constraints (\ref{pure-constr}) reduce the dimension of the
$\vec{n}$-space to four.\cite{Ercolessi+01,LoktevOstrovskii94}

For a mixed state, Eq.\ (\ref{dmatrix-n}) retains sense as it is
the most general expression for a $3\times 3$ matrix with unit
trace. The constraints (\ref{pure-constr}) generally do not hold for a
mixed state. However, under a unitary rotation
$\widehat{\rho}(\vec{n})\mapsto
U^{\dag}\widehat{\rho}(\vec{n})U=\widehat{\rho}(\vec{n'})$ the
transformation $\vec{n}\mapsto\vec{n'}$ is still determined by the
adjoint representation of SU(3), so
$I_{2,3}(\vec{n})=I_{2,3}(\vec{n'})$ remain invariant, but are no more
fixed at the pure state values (\ref{pure-constr}). The dimension of
the $\vec{n}$-space in the general case of a mixed state
(\ref{dmatrix-n}) is equal to six.\cite{Ercolessi+01,LoktevOstrovskii94}

The components of the octet $\vec{n}$ correspond to
the following on-site spin averages:
\begin{eqnarray}
\label{8-aver}
&&n_{2}=\langle S^{z}\rangle, \quad
 n_{5}=-\langle S^{y}\rangle,\quad
 n_{7}=\langle S^{x}\rangle, \nonumber\\
&& n_{1}=\langle S^{x}S^{y}+S^{y}S^{x}\rangle, \quad n_{4}=-\langle S^{x}S^{z}+S^{z}S^{x}\rangle, \nonumber\\
&& n_{6}=\langle S^{y}S^{z}+S^{z}S^{y}\rangle,\quad  \\
&& n_{3}=\langle (S^{x})^{2}-(S^{y})^{2}\rangle,\quad
n_{8}=\sqrt{3}\big(\langle (S^{z})^{2}\rangle -2/3\big),\nonumber
 \end{eqnarray}
which can be split into the magnetization (dipole) part $\vec{m}$
and nematic (quadrupolar) one $\vec{d}$,
\begin{eqnarray}
\label{md}
&&\vec{n}=\vec{m}+\vec{d},\nonumber\\
&& \vec{m}=(n_{7},-n_{5},n_{2}),\quad
\vec{d}= (n_{1},n_{3},n_{4},n_{6},n_{8}).
\end{eqnarray}
It is easy to see that under time reversal operation $\widehat{T}$, the magnetization $\vec{m}$
changes sign, while the quadrupolar part $\vec{d}$ remains invariant:
\begin{equation}
\label{Tm-Td}
\widehat{T}\vec{m}=-\vec{m},\quad \widehat{T}\vec{d}=\vec{d}.
\end{equation}

\subsection{Equations of motion for pure spin $S=1$ states}

The effective Lagrangian of a spin-1 system on a lattice, expressed in terms of the complex unit
vector $\vec{z}$, takes the form
\begin{equation}
\label{lagr-cp2d}
\mathcal{L}=\sum_{j}i(\vec{z}_{j}^{*}\cdot
\partial_{t} \vec{z}_{j})-W(\{\vec{z}_{j}^{*},\vec{z}_{j}\}),
\end{equation}
where $W$ is the  energy that can be also
expressed through the octets $\{\vec{n}_{j}\}$. We will assume that
one can pass to the continuum description just by declaring $\vec{z}$
(or, alternatively, $\vec{n}$)
a smooth field. Then the
Lagrange equations of motion for $\vec{z}^{*}$, $\vec{z}$ take the form
\begin{equation}
\label{eqm-z}
i\frac{\partial\vec{z}^{*}}{\partial t}=-\frac{\delta
  W}{\delta\vec{z}},
\end{equation}
where the energy $W[\vec{z}^{*},\vec{z}]=W[\vec{n}]$ is now a
functional of either the complex vector field $\vec{z}$ or the real
octet field $\vec{n}$, and $\delta/\delta\vec{z}$ denotes a
variational derivative.
Introducing generalized fields
\begin{equation}
\label{gen-field}
\vec{H}=-\frac{\delta W}{\delta \vec{n}},\quad \vec{h}=-\frac{\delta  W}{\delta \vec{z}},
\end{equation}
one can write
\begin{eqnarray*}
&& \frac{\partial n_{\alpha}}{\partial t} =-i
  (h_{a}\lambda^{\alpha}_{ab}z_{b}- \mbox{c.c}),\\
&& h_{b}=H_{\alpha}\lambda^{\alpha}_{ab}z^{*}_{a},
\end{eqnarray*}
which finally leads to the equation of motion for $\vec{n}$:
\begin{equation}
\label{eqm-n}
\frac{\partial \vec{n}}{\partial t}=2 (\vec{n}\wedge \vec{H}).
\end{equation}
It is worth noting that
if $W$ depends only on magnetization $\vec{m}$, (\ref{eqm-n}) reduces
to the well-known Landau-Lifshitz equation
without dissipation,
$\partial\vec{m}/\partial t= -(\vec{m}\times \delta W/\delta
\vec{m})$, which describes spin dynamics in ferromagnets.
Thus, (\ref{eqm-n}) can be viewed as
an extension of the
Landau-Lifshitz equation that includes the dynamics of quadrupolar
degrees of freedom.

\subsection{General form of the equations of motion and relaxation terms}

General analysis of dissipative forces in spin-1 nematics can be
carried out closely following the approach  developed for ferromagnets.~\cite{Baryakhtar84} Equations of motion  can generally be cast in the form of Onsager equations
\begin{equation}
\label{Onsager-eqs}
\frac{\partial n_{\alpha}}{\partial t}=\widehat{\Lambda}_{\alpha\beta}(\vec{n}) H_{\beta},
\end{equation}
where  the effective fields $\vec{H}$, defined in (\ref{gen-field}),
play the role of generalized forces that arise at a deviation from equilibrium,
and $\widehat{\Lambda}_{\alpha\beta}$ are the kinetic coefficients
that generally shall be understood as operators acting on $\vec{H}$
(physically this corresponds to taking into account the spatial
dispersion).

Since different components of the octet vector $\vec{n}$ have
different properties under time reversal, the Onsager
theorem about the symmetry of the kinetic coefficients
$\Lambda_{\alpha\beta}(\vec{n})$ gets slightly modified.\cite{Onuki-book} If one
introduces the factors $\epsilon_{\alpha}=\pm 1$ describing the symmetry
of the $\alpha$-th component of $\vec{n}$  under time reversal,
\begin{equation}
\label{Tna}
\widehat{T} n_{\alpha}=\epsilon_{\alpha}n_{\alpha},
\end{equation}
then the kinetic coefficients  satisfy
the following reciprocity relations:
\begin{equation}
\label{Onsager-rels}
\widehat{\Lambda}_{\beta\alpha}(\vec{n})=\epsilon_{\alpha}\epsilon_{\beta}\widehat{\Lambda}_{\alpha\beta}(\widehat{T}\vec{n}).
\end{equation}

It is easy to see that only the symmetric part
$\widehat{\Lambda}^{(s)}_{\alpha\beta}$ of the tensor
$\widehat{\Lambda}$ contributes to relaxation. Indeed,
the dissipative function $Q$ can be written as
\begin{eqnarray}
\label{diss-fun}
Q&=&-\frac{1}{2}\frac{dW}{dt}=\frac{1}{2}\int d\vec{x} \, \vec{H}\cdot
\frac{\partial\vec{n}}{\partial t}\nonumber\\
&=&\frac{1}{2}\int d\vec{x}\,
H_{\alpha}\widehat{\Lambda}_{\alpha\beta}H_{\beta}
= \frac{1}{2}\int d\vec{x}\,
H_{\alpha}\widehat{\Lambda}^{(s)}_{\alpha\beta}H_{\beta}.
\end{eqnarray}
The antisymmetric part $\widehat{\Lambda}^{(a)}_{\alpha\beta}$ of the
kinetic coefficients tensor makes no contribution into dissipation and thus
corresponds to the purely dynamic part of the equations of
motion. As we have already established, those
equations are given by  (\ref{eqm-n}), and thus one obtains
\begin{equation}
\label{Lambda-asym}
\widehat{\Lambda}^{(a)}_{\alpha\beta}(\vec{n}) =-2f_{\alpha\beta\gamma}n_{\gamma}.
\end{equation}
Under the assumption of weak dispersion, one can expand the symmetric part of $\widehat{\Lambda}$
in spatial derivatives, retaining only up to quadratic terms.

Assuming further that the crystal structure of the magnet has an inversion center, one can ignore the term linear in derivatives, so
finally we obtain:
\begin{equation}
\label{Lambda-sym}
\widehat{\Lambda}^{(s)}_{\alpha\beta}(\vec{n})=\lambda_{\alpha\beta}(\vec{n})
-\widetilde{\lambda}_{\alpha\beta,ll'}(\vec{n})
\frac{\partial^{2}}{\partial x_{l} \,\partial x_{l'}} +\ldots,
\end{equation}
where $\lambda_{\alpha\beta}$ and
$\widetilde{\lambda}_{\alpha\beta,ll'}$
satisfy the reciprocity relations similar to (\ref{Onsager-rels}) with
respect to permutations of $\alpha$ and $\beta$, and $\widetilde{\lambda}_{\alpha\beta,ll'}$
are symmetric with respect to permutations of $l$ and $l'$. The dissipation function in this approximation will contain
terms quadratic in the effective fields and their derivatives,
\begin{equation}
\label{diss-fun2}
Q=\int d\vec{x}\, \Big\{
\lambda_{\alpha\beta}H_{\alpha}H_{\beta}+\widetilde{\lambda}_{\alpha\beta,ll'}
\frac{\partial H_{\alpha}}{\partial x_{l}}\frac{\partial
  H_{\beta}}{\partial x_{l'}} \Big\},
\end{equation}
and equations of motion including relaxation terms acquire the
following general form:
\begin{eqnarray}
\label{eqm-gen}
\frac{\partial n_{\alpha}}{\partial t}&=&2f_{\alpha\beta\gamma}
n_{\beta}H_{\gamma}+R_{\alpha} \nonumber\\
R_{\alpha}&=&\frac{\delta Q}{\delta
  H_{\alpha}}=\lambda_{\alpha\beta}H_{\beta}-\widetilde{\lambda}_{\alpha\beta,ll'}
\frac{\partial^{2}H_{\beta}}{\partial
x_{l}\, \partial x_{l'}}.
\end{eqnarray}
Similarly as it is done for usual ferromagnets,\cite{spinwaves,Baryakhtar84} at finite
  temperatures the energy $W$ in Eq.\ (\ref{lagr-cp2d}) has to be
  understood as the free energy, and the entire framework has to be
  considered as a phenomenological time-dependent Ginzburg-Landau
  theory, with the parameters of the free energy and the dissipative
  function being some temperature-dependent constants.

\section{Phenomenology applied to bilinear-biquadratic model}

We apply the general formalism presented above to the
bilinear-biquadratic model described by the Hamiltonian
\begin{equation}
\label{ham-bilbiq}
\widehat{\mathcal{H}}=-J_{1} \sum_{\langle
  ij\rangle}\vec{S}_{i}\cdot\vec{S}_{j} -J_{2}\sum_{\langle
  ij\rangle}(\vec{S}_{i}\cdot\vec{S}_{j})^{2},
\end{equation}
where $\vec{S}_{i}$ are spin-1 operators at the $i$-th lattice site, and the
sums are over nearest neighbors. For simplicity, we assume that the
lattice is cubic, and the
lattice constant is set to unity.

The above model describes the most general isotropic ($SU(2)$ invariant) exchange interaction between two $S=1$ spins.
At $J_{1}=J_{2}$ the symmetry of the model is enhanced to $SU(3)$.
We consider the nematic region in the vicinity  of the
ferromagnetic $SU(3)$ point, so it is convenient to set
\begin{equation}
\label{Js-ferronem}
J_{1}=J(1-\delta),\quad J_{2}=J,\quad J>0,\quad 0<\delta<1/2.
\end{equation}
Introducing the continuum field $\vec{n}=\vec{m}+\vec{d}$, one can
write down the energy $W=\langle \widehat{\mathcal{H}} \rangle$ as
\begin{eqnarray}
\label{W-bilbiq}
W&=&\int d^{3}x\, \Big\{ (Z/2)\big[ -\frac{1}{2}J\vec{n}^{2}+
\delta J\vec{m}^{2}\big] \nonumber\\
&+&\frac{J}{4}\big[ (\nabla \vec{d})^{2} +(1-2\delta) (\nabla \vec{m})^{2}\big]
 \Big\},
\end{eqnarray}
where $Z=6$ is the lattice coordination number, and $(\nabla \vec{n})^{2}\equiv
\sum_{l}(\partial\vec{n}/\partial x_{l})^{2}$.  For pure states, the
constraints (\ref{pure-constr}) fix the length of $\vec{n}$; thus, for
$\delta>0$, in the ground state the magnetization $\vec{m}$ vanishes,
which corresponds to a uniaxial spin nematic.

At finite temperature $T$, expression (\ref{W-bilbiq}) has to be
replaced by the free energy, $\mathcal{F}=W-T\mathcal{S}$, where
$\mathcal{S}$ is the entropy of the system. Below we use a
phenomenological expression for the free energy written in the spirit
of Landau's theory of phase transitions in a form of expansion on powers
of order parameters. This expression generally can be constructed from
all possible invariants of the corresponding symmetry group.

We limit ourselves to
the case $\delta<1/2$, when local spin correlations are ferromagnetic.
For $\delta>1/2$ the nearest-neighbor spin-spin correlations change
their character to antiferromagnetic, and the effective continuum theory has to be
modified ($\vec{n}$ has to be split into a uniform and staggered
components that become smooth fields in the continuum theory \cite{Kolezhuk08}). The
detailed discussion of this interesting case is going far beyond the
scope of this article, and below we will limit ourselves to the case
$ \delta <1/2 $, paying main attention to the vicinity of
ferromagnetic $SU(3)$ point, where $ \delta \ll 1$.

As a consequence of the assumed cubic symmetry of the lattice, the
tensor of dissipative constants, that enters relaxation forces
depending on derivatives, must be diagonal in its space indices,
\[
\widetilde{\lambda}_{\alpha\beta,ll'}=\widetilde{\lambda}_{\alpha\beta}\delta_{ll'}.
\]
 Now let
us focus separately on the highly symmetric case $\delta=0$ and then
study what happens at deviations from this point.

\subsection{$SU(3)$ symmetric case}

At $\delta=0$ the system is $SU(3)$-symmetric, which
dictates that at this point both tensors of dissipative constants are
diagonal in the octet indices, $\lambda_{\alpha\beta}=\lambda\delta_{\alpha\beta}$ and
$\widetilde{\lambda}_{\alpha\beta}=\widetilde{\lambda}\delta_{\alpha\beta}$. The
dissipative force $\vec{R}$ is thus determined by just two constants
$\lambda$ and $\widetilde{\lambda}$:
\begin{equation}
\label{R-su3}
\vec{R}=\lambda \vec{H}-\widetilde{\lambda}\, \nabla^{2}\vec{H}.
\end{equation}
The free energy $\mathcal{F}$ will generally, in addition to the gradient term, contain some function
$f(I_{2},I_{3})$ of
the $SU(3)$ invariants $I_{2}(\vec{n})$ and $I_{3}(\vec{n})$, that
will have a minimum at certain equilibrium values of $I_{2}(\vec{n}_{0})=I_{2}^{(0)}$ and
$I_{3}(\vec{n}_{0})=I_{3}^{(0)}$. Keeping only the lowest order terms in the
expansion of $f(I_{2},I_{3})$ around its minimum as well as in the
gradient expansion,  we obtain
\begin{eqnarray}
\label{W-su3}
\mathcal{F}&=& \int d^{3}x\, \Big\{ \frac{(\vec{n}^{2}-I_{2}^{(0)})^{2}}{4\chi_{2}} +
\frac{(\vec{n}\cdot(\vec{n}*\vec{n})-I_{3}^{(0)})^{2}}{6\chi_{3}}  \nonumber\\
&+&\frac{\widetilde{J}}{4} (\nabla \vec{n})^{2}
 \Big\},
\end{eqnarray}
where $\chi_{2}$ and $\chi_{3}$ play the role of longitudinal
susceptibilities.
In the above expression for the free energy, all parameters have to be
understood as phenomenological constants that are generally some functions of
the temperature.
At low temperatures $T\ll J$, the temperature dependence of the 
effective exchange parameter
$\widetilde{J}$ is rather weak,  so in what follows
we assume that $\widetilde{J}\approx J$.

The effective field $\vec{H}=-\delta \mathcal{F}/\delta \vec{n}$ takes the
form
\begin{eqnarray}
\label{Heff-su3}
\vec{H}&=&-\big[\big(\vec{n}^{2}-I_{2}^{(0)}\big)/\chi_{2}\big] \vec{n} \\
&-&\big[\big(\vec{n}\cdot(\vec{n}*\vec{n})-I_{3}^{(0)}\big)/\chi_{3}\big]
(\vec{n}*\vec{n})
+\frac{J}{2}\nabla^{2}\vec{n},\nonumber
\end{eqnarray}
where the first two terms do not contribute to the equation of motion (\ref{eqm-n})
in absence of dissipation  (note that $\vec{n}\wedge (\vec{n}*\vec{n})=0$).

Further, at $\delta=0$ the quantity $\vec{N}=\int d^{3}x\, \vec{n}$ is
conserved. As a consequence, the equation of motion for $\vec{n}$
should have the form of a continuity equation $\partial
n_{\alpha}/\partial t+ \mbox{div} \vec{\Pi}_{\alpha}=0$ and so the
dissipative force $\vec{R}$ must have the form $R_{\alpha}=\partial
\Pi_{\alpha b}/\partial x_{b}$. It is clear from (\ref{R-su3}) and
(\ref{Heff-su3}) that $\vec{R}$ can be represented in such a form only
if $\lambda=0$. \emph{Thus, at the $SU(3)$ symmetric point $\delta=0$
  we are left with a single relaxation constant
  $\widetilde{\lambda}$:}
\begin{equation}
\label{R-su3a}
\vec{R}=-\widetilde{\lambda}\, \nabla^{2}\vec{H}.
\end{equation}
The above argument is fully similar to the treatment of the exchange approximation in ferromagnets.\cite{Baryakhtar84}

Let us calculate the magnon damping in spin nematic at the
$SU(3)$-symmetric point. Without  loss of generality, the uniform ground state
can  be chosen as $\vec{n}=\vec{n}_{0}$ with
\begin{equation}
\label{n0}
\vec{n}_{0}=n_{0}(0,0,0,0,0,0,0,-1) ,
\end{equation}
where $n_{0}\equiv\sqrt{I_{2}^{(0)}}$ is also temperature-dependent.
This state corresponds to a uniaxial spin nematic with zero
magnetization and the nematic director along the $z$-axis. To
visualize it, one may think of an ellipsoid with the same symmetry as
for the set of quadrupole averages $\langle S_iS_j+S_jS_i\rangle$. For
the above state, such an image is a squeezed ellipsoid of rotation,
with the ellipsoid axis being directed along the $z$ axis, and with
the thickness $\propto \langle S_z^2\rangle $ smaller than the
diameter $\propto \langle S_x^2\rangle=\langle S_y^2\rangle$.  At zero
temperature $n_{0}=2/\sqrt{3}$ and thus $\langle S_z^2\rangle = 0$,
i.e., the ellipsoid degenerates into a zero thickness disk.  Due to
the high symmetry of the system, any other choice of $\vec{n}_{0}$
that can be obtained from (\ref{n0}) by a general $SU(3)$ rotation
would give the same physical results. We have chosen the ground state
(\ref{n0}) because it has the simplest form in the octet
representation.

We consider small deviations from the ground state,
$\vec{n}=\vec{n}_{0}+\vec{\eta}$,  linearize equations of motion
(\ref{eqm-gen}) in $\vec{\eta}$, and look for eigenmodes in the form
$\vec{\eta}=\widetilde{\vec{\eta}}e^{i(\vec{k}\cdot\vec{x}-\omega
  t)}$.
Then the damping $\Gamma_{k}$ is obtained  as the imaginary
part of the frequency $\omega_{k}=\Omega_{k}-i\Gamma_{k}$. Since we
are interested only in the case of weak inhomogeneities, one can
expand $\Omega_{k}$ and $\Gamma_{k}$ in powers of the wave vector
$\vec{k}$ and retain only leading terms.

It should be emphasized that in the present work we focus on
  the behavior of three-dimensional nematics. In three dimensions, the
  ordered ground state
(\ref{n0}) with spontaneously broken symmetry survives up to a certain
 temperature $T_{c}$, and from numerical simulations the value of
the critical temperature is
known \cite{HaradaKawashima2002} to be sufficiently high: around the
$SU(3)$ point $T_{c}$ is roughly equal to the microscopic exchange
constant $J$. In low-dimensional systems, thermal or quantum
fluctuations can destroy the long-range order, making the above
approach inapplicable.

It turns out that the dynamics of the components $\eta_{1}$,
$\eta_{2}$, $\eta_{3}$, and $\eta_{8}$ is purely diffusive: the real
parts $\Omega_{k}$ of the  corresponding frequencies vanish, and the
damping behavior is  given by
\begin{eqnarray}
\label{Freq-su3-diff}
&& \Gamma_{1}=\Gamma_{2}=\Gamma_{3}=\frac{1}{2}\widetilde{\lambda}J k^{4},\\
&& \Gamma_{8}= C\widetilde{\lambda}k^{2}+ \frac{1}{2}\widetilde{\lambda}J
k^{4},\quad C=n_{0}^{2}\Big(\frac{2}{\chi_{2}}+\frac{n_{0}^{2}}{\chi_{3}} \Big).\nonumber
\end{eqnarray}
The above  modes are longitudinal, they correspond to  hydrodynamic relaxation
of inhomogeneities in the distribution of  $\vec{n}$ and are not
connected to any propagating excitations.
Let us briefly discuss their physical meaning. The variable
$\eta_{2}=m_z$ describes the $z$ component of the magnetization,
and two variables
$\eta_{1}$, $\eta_{3}$ correspond to
quadrupolar averages built from projections of the spin in the $(xy)$ plane.
The dynamics of these variables can be viewed as a change of the spin length
 coupled with the rotation of the system around
the $z$ axis. Such a geometry corresponds to the longitudinal mode
found in the same model at $\delta<0$ (i.e.,  in the ferromagnetic state),
where this mode acquires a finite frequency and thus becomes
propagating.\cite{Ivanov+prb08}
The
last diffusive mode $\eta_{8}$ determines the evolution of the variable $\langle
m_z^2 \rangle$ and describes the relaxation of the de Gennes order
parameter to its equilibrium value. In the $SU(3)$-symmetric case
there is only inhomogeneous relaxation (all damping constants contain
$k$). As we shall see below, a deviation from this high symmetry point
leads to the appearance of a homogeneous relaxation for quadrupolar
degrees of freedom, but not for the total magnetic moment.

The rest of the modes are dynamical and correspond to propagating
generalized magnons that unite dipolar (spin) and quadrupolar excitations.
The dynamics of $\eta_{4}$ and $\eta_{5}$ is coupled, and the same
holds for the pair $\eta_{6}$, $\eta_{7}$. The corresponding
eigenfrequencies are
\begin{eqnarray}
\label{Freq-su3-dyn}
&& \omega_{(45)}=\pm \Omega_{(45)}-i\Gamma_{(45)}, \quad \omega_{(67)}=\pm \Omega_{(67)}-i\Gamma_{(67)},
\\
&& \Omega_{(45)}=\Omega_{(67)}=\frac{Jn_{0}\sqrt{3}}{2}k^{2},\quad
\Gamma_{(45)}=\Gamma_{(67)}=\frac{1}{2}\widetilde{\lambda}J k^{4}.\nonumber
\end{eqnarray}
One can see that at small wavevectors the magnon damping behaves as
$\Gamma_{k}\propto k^{4}$, while the real
part of the frequency $\Omega_{k}\propto k^{2} \gg \Gamma_{k}$, so
magnons remain well-defined excitations.

\subsection{Spin nematic away from the $SU(3)$ point}

Consider a system in the nematic phase ($1/2>\delta>0$). In the
ground state the magnetization $\vec{m}$ vanishes, so the equilibrium
value $\vec{n}=\vec{n}_{0}\equiv\vec{d}_{0}$ lies completely within
the quadrupolar subspace. The
symmetry group is now reduced to $SU(2)$ which is a subgroup of $SU(3)$. Under  rotations,
$\vec{m}$ and $\vec{d}$ transform according to $D^{1}$ and $D^{2}$
representations of the $SU(2)$ group, respectively. Thus,
$\vec{m}^{2}$ and $\vec{d}^{2}$ are invariants that can separately enter the free
energy. The $SU(3)$ invariants $I_{2}(\vec{n})$ and $I_{3}(\vec{n})$,
of course, remain invariant under any subgroup transformation; since
$I_{2}=\vec{m}^{2}+\vec{d}^{2}$, we shall only take into account
$I_{3}(\vec{n})$ as another independent invariant.
We will assume that the free energy has the form
\begin{eqnarray}
\label{W-su2}
\mathcal{F}&=& \int d^{3}x\, \Big\{ \frac{Z}{2}\delta J\vec{m}^{2}+
\frac{(\vec{d}^{2}-\vec{n}_{0}^{2})^{2}}{4\chi_{d}} \nonumber\\
&+&
\frac{\big[ \vec{n}\cdot(\vec{n}*\vec{n})-I_{3}(\vec{n}_{0}) \big]^{2}}{6\chi_{3}}  \nonumber\\
&+&\frac{J}{4}\big[ (\nabla \vec{d})^{2} +(1-2\delta)(\nabla \vec{m})^{2} \big]
 \Big\},
\end{eqnarray}
where we have neglected the terms of higher than quadratic order in
$\vec{m}$ and $(\vec{d}-\vec{n}_{0})$.

Similar to (\ref{md}), we can divide the effective field $\vec{H}$
into two components $\vec{H}_{m}$ and $\vec{H}_{d}$ lying in the
dipolar and quadrupolar subspaces of the octet space, and transforming
according to $D^{1}$ and $D^{2}$ representations of the $SU(2)$ group,
respectively. Then, the symmetry dictates the following structure of
the dissipative constants tensors:
\begin{equation}
\label{R-su2}
\vec{R}=0\cdot\vec{H}_{m}-\widetilde{\lambda}_{m}\nabla^{2}\vec{H}_{m}
+\lambda_{d}\vec{H}_{d}-\widetilde{\lambda}_{d}\nabla^{2}\vec{H}_{d},
\end{equation}
where the vanishing constant in front of $\vec{H}_{m}$ is the
consequence of the fact that the total magnetization $\vec{M}=\int
d^{3}x\, \vec{m}$ is an integral of motion. From our
analysis of the $SU(3)$-symmetric point it also follows that
\begin{equation}
\label{su2-su3}
 \lambda_{d}\to 0,\quad
\widetilde{\lambda}_{m}\to\widetilde{\lambda}_{d} \quad \mbox{at}\quad
\delta\to 0.
\end{equation}
Eqs.\ (\ref{R-su2},\ref{su2-su3}) are valid on both sides
of the $SU(3)$ point (i.e., at any sign of $\delta$), while
$\lambda_{d}$ is strictly non-negative. If one assumes
$\lambda_{d}$ to behave analytically as the function of $\delta$,
then it follows that
\begin{equation}
\label{lambda-d}
\lambda_{d}=O(\delta^{2}) \quad \mbox{at}\quad
\delta\to 0,
\end{equation}
however, on general grounds one cannot exclude the possibility of
singular behavior of $\lambda_{d}(\delta)$. We will check the above
assumption and its consequence (\ref{lambda-d}) in the next section by
comparing it with the results of microscopic analysis.

Let us now calculate the magnon damping away from the $SU(3)$ point.
Without loss of generality, we assume the nematic ground state with
the director along the $z$ axis, i.e., $\vec{n}_{0}$ of the form
(\ref{n0}). We linearize equations of motion in the deviation
$\vec{\eta}=\vec{n}-\vec{n}_{0}$ and find the complex eigenfrequencies
$\omega=\Omega-i\Gamma$, as we have done before in the $SU(3)$
case. We find that there are again four longitudinal, purely diffusive
($\Omega=0$), decoupled modes $\eta_{1}$, $\eta_{2}$, $\eta_{3}$, and
$\eta_{8}$, with the following linewidths:
\begin{eqnarray}
\label{Freq-su2-diff}
&& \Gamma_{1}=\Gamma_{3}
=\frac{1}{2}J k^{2}(\lambda_{d}+\widetilde{\lambda}_{d}k^{2}),\nonumber\\
&& \Gamma_{2}=\widetilde{\lambda}_{m} J k^{2}\big[Z\delta +\frac{1}{2}(1-2\delta)k^{2}\big],\\
&& \Gamma_{8}= \Big(C +\frac{1}{2}J
k^{2}\Big)(\lambda_{d}+\widetilde{\lambda}_{d}k^{2}),\quad
C=n_{0}^{2}\Big(\frac{2}{\chi_{d}}+\frac{n_{0}^{2}}{\chi_{3}} \Big).\nonumber
\end{eqnarray}
Comparing the above expressions with their $SU(3)$ counterparts in
Eqs.\ (\ref{Freq-su3-diff}), one can see that the main effect of
breaking the $SU(3)$ symmetry is the appearance of a homogeneous
relaxation in the $\eta_{8}$ quadrupolar order parameter proportional
to $\langle(S^z)^2-2/3\rangle$: $k=0$ fluctuations of $\langle
S_{z}^{2}\rangle$ decay with the characteristic relaxation time
\begin{equation}
\label{diff-tau}
\tau_{8}\equiv \frac{1}{\Gamma_{8}(k=0)}=\frac{1}{C\lambda_{d}},
\end{equation}
which diverges  at $\delta\to0$ according to Eq.\ (\ref{lambda-d}).

As before, there is no homogeneous relaxation of the magnetization,
because the $SU(2)$ symmetry remains intact and the total
magnetization is conserved. For the other three diffusive modes
$\eta_{1,2,3}$, there is a small (of the order of $\delta$) splitting
between the damping constants of the quadrupolar modes ($\eta_{1,3}$)
and the magnetization mode ($\eta_{2}$), and the leading term in the
damping at small wave vectors behaves as $k^{2}$ (in contrast to the
$k^{4}$ behavior in the $SU(3)$ case).

The remaining  two degenerate eigenmodes
correspond to coupled  $(\eta_{4},\eta_{5})$ and
$(\eta_{6},\eta_{7})$ oscillations,
and describe propagating magnons. The corresponding
eigenfrequencies $\Omega_{(45)}=\Omega_{(67)}$  and dampings $\Gamma_{(45)}=\Gamma_{(67)}$ are given by
\begin{eqnarray}
\label{Freq-su2-dyn}
\Omega_{(45)}=\Omega_{(67)}&=&\frac{Jn_{0}\sqrt{3}}{2}|k|\Big(2Z\delta+(1-2\delta)k^{2}
\Big)^{1/2},\nonumber\\
 \Gamma_{(45)}=\Gamma_{(67)}&=&
\frac{Jk^{2}}{4}\Big\{ \lambda_{d}+2Z\delta\widetilde{\lambda}_{m}\\
&+&
\big[\widetilde{\lambda}_{d}+(1-2\delta)\widetilde{\lambda}_{m}\big]k^{2}\Big\} .\nonumber
\end{eqnarray}
It is easy to see that away from the $SU(3)$ point magnons acquire
linear dispersion, in agreement with previous results obtained by
different
methods,\cite{Papanicolaou1988,Chubukov1990,IvanovKolezhuk03} and the
leading term in their damping at small wavevectors behaves as
$k^{2}$. At $\delta\to 0$, taking into account (\ref{su2-su3}), we
recover our results obtained above for the $SU(3)$ point.

\section{Microscopic analysis of the bilinear-biquadratic model}
\label{sec:micro}

Our goal is to compare the results for magnon damping in the
bilinear-biquadratic model (\ref{ham-bilbiq}), obtained above within
the phenomenological approach, with a calculation from first
principles.  We would like to pass to the second-quantization
formalism for the description of magnons. It is convenient to break up
the complex vector $\vec{z}$, describing the coherent state
(\ref{states}), into two real vectors representing its real and
imaginary parts $\vec{z}=\vec{u}+i\vec{v}$. When the normalization
condition and the arbitrariness of the phase factor are taken into
account, these vectors satisfy the conditions:
\begin{equation}
\label{uv-constr}
\vec{u}^2+\vec{v}^2=1, \quad \vec{u}\cdot\vec{v}=0.
\end{equation}
In terms of the variable vectors $\vec{u}$ and $\vec{v}$ the Lagrangian of the system can be written as
\begin{eqnarray}
\mathcal{L}&=&-2\hbar \sum_{j}
\vec{v}_{j}\partial_{t}\vec{u}_{j}-W,\nonumber\\
 W&=&\sum_{\langle ij\rangle} \Big\{ 2(J_2-J_1)\left[(\vec{u}_i\vec{u}_j)(\vec{v}_i\vec{v}_j)-
(\vec{u}_i\vec{v}_j)(\vec{v}_i\vec{u}_j)\right]\nonumber\\
&&-\frac{J_2}{2}\left[(\vec{u}_i\vec{u}_j+\vec{v}_i\vec{v}_j)^2+
(\vec{u}_i\vec{v}_j-\vec{v}_i\vec{u}_j)^2\right]\Big\}.
\end{eqnarray}

The spin nematic state, favored at
$J_2>J_1>0$,  corresponds to the following condition: vectors
$\vec{u}_{j} $ at all the sites are parallel,
$\vec{u}_{j} =\vec{u}_0 $, $| \vec{u}_0 | =1$, and
$\vec{v}_{j} =0$. (There is also an alternative configuration
obtained by the substitutions $\vec{u}_{j} \mapsto
\vec{v}_{j} ,\vec{v}_{j} \mapsto -\vec{u}_{j} $, but
those two states are physically identical).

Consider small deviations of the variables $\vec{u}$ and
$\vec{v}$ from the spin nematic state. Assume, for the sake of
definiteness, that $\vec{u}_{0}=\vec{e}_{z}$. Then, according to
(\ref{uv-constr}), $v_{z} $ and $u_{z} $ are dependent variables quadratic in
the remaining components $u_{x,y}$ and $v_{x,y}$.  We can thus expand the Lagrangian in powers of the
independent variables $u_{x,y}$ and $v_{x,y}$, keeping terms up to the
quartic order.
From the kinetic part of the Lagrangian it is obvious that $u_{x}$ and
$u_{y} $ can be chosen as the coordinates, while $-2\hbar v_{x}
$ and $-2\hbar v_{y} $ play the role of the corresponding
canonical momenta. It is easy then to pass from the Lagrangian to the
Hamiltonian.
 Passing from variables $\vec{u}_{j}$,
$\vec{v}_{j}$ to their Fourier amplitudes
$\widetilde{\vec{u}}_{\vec{k}}$, $\widetilde{\vec{v}}_{\vec{k}}$ and
 performing quantization,
we can write the standard representation of the
coordinates and momenta in terms of the Bose creation and
annihilation operators:
\begin{eqnarray}
\label{2nd-q}
&&\widetilde{u}_{x,\vec{k}} =\sqrt{\frac{\hbar }{2A_{\vec{k}} }} (a_{\vec{k}}^\dag +a_{-\vec{k}} ), \;
\widetilde{v}_{x,\vec{k}} =i\sqrt{\frac{A_{\vec{k}} }{8\hbar }} ( a_{\vec{k}}^\dag -a_{-\vec{k}}),\nonumber\\
&&\widetilde{u}_{y,\vec{k}} =
\sqrt{\frac{\hbar }{2A_{\vec{k}} }}
(b_{\vec{k}}^\dag +b_{-\vec{k}}), \;
\widetilde{v}_{y,\vec{k}} =i\sqrt{\frac{A_{\vec{k}} }{8\hbar }}
(b_{\vec{k}}^\dag -b^{\vphantom{\dag}}_{-\vec{k}}),
\end{eqnarray}
where
$A_{\vec{k}} =2\hbar \sqrt {\alpha _{\vec{k}} /\beta_{\vec{k}} }$,
and we use the notation
\begin{equation}
\label{alpha-gamma}
\begin{split}
 \alpha_{\vec{k}} =ZJ_2 \left[ {1-\gamma( \vec{k} )} \right],\;\;
 \beta_{\vec{k}} =J_2 k_{0}^{2} \gamma( \vec{k})+\alpha_{\vec{k}} , \\
 k_{0}^{2}= 2Z( 1-J_1 /J_2 )\equiv 2Z\delta, \;\; \gamma(\vec{k})=\frac{1}{Z}\sum_{\vec\ell  } {e^{i\vec k\vec \ell
}} ,
\end{split}
\end{equation}
$\vec \ell$ being the set of vectors connecting a site of the lattice
to its nearest neighbors.

The Hamiltonian
decomposes into a sum of quadratic and quartic terms:
$\widehat{H}=\widehat{H}_2 +\widehat{H}_4$.
The quadratic part $H_2 $ becomes diagonal,
\begin{equation}
\label{H2}
H_2 =\sum_{\vec{k}} \varepsilon _{\vec{k}} ( a_{\vec{k}}^\dag a_{\vec{k}}
+b_{\vec{k}}^\dag b_{\vec{k}}  ) ,
\end{equation}
and the  magnon dispersion relation reads
\begin{equation}
\label{epsilon}
\varepsilon _{\vec{k}} =\sqrt {\alpha_{\vec{k}} \beta_{\vec{k}} } .
\end{equation}

The spectrum contains two degenerate magnon modes with orthogonal
polarizations, that are gapless at $k\to 0$ in accordance to
the Goldstone theorem. For $J_1 >J_2 $ this spectrum becomes unstable, which corresponds
to a transition to the ferromagnetic state.

In this paper, we are interested in the low-energy
dynamics. In the vicinity of the ferronematic $SU(3)$ point
$J_{1}=J_{2}$ (i.e., $\delta=0$) this is translated into the long-wavelength
approximation $k \ll 1$, so we obtain $\gamma(\vec{k})\approx 1-k^{2}/Z$
and
\begin{equation}
\label{kll1}
\varepsilon_{\vec{k}} = J_2 k \sqrt {k_{0}^{2} +k^{2}} \quad \text{at
  $k\ll 1$}.
\end{equation}

The quartic Hamiltonian $\widehat{H}_{4}$ is rather cumbersome, but
for our purpose of calculating the magnon damping in the lowest order
of the perturbation theory it can be simplified substantially. Indeed,
the analysis shows that decay processes (one magnon decaying into
three) give a small contribution at low energies, see the note
after Eq.(\ref{highT}) below. Thus, the main contribution to the
damping comes from terms containing an equal number of creation and
annihilation operators. When only such terms are taken into account,
the Hamiltonian takes the form
\begin{eqnarray}
\label{H4}
\widehat{H}_4 &=&\frac{1}{N}\sum_{1,2,3,4} \Delta_{1+2-3-4}
\Big\{ \Phi a_1^\dag a_2^\dag b_3 b_4   +\Psi a_1^\dag b_2^\dag a_3 b_4
\nonumber \\
&+&F( a_1^\dag a_2^\dag a_3 a_4 +b_1^\dag b_2^\dag b_3 b_4 )+\text{h.c.}  \Big\},
\end{eqnarray}
where for the sake of brevity we use the shorthand notation $1\equiv
\vec{k}_{1}$, etc.

The amplitudes $\Phi$, $\Psi$, and $F$ in Eq.\ (\ref{H4}) depend on
all four magnon momenta. $\Phi$ corresponds to processes of conversion
between pairs of magnons with different polarizations, while the other
two amplitudes describe magnon scattering. Further, in our case (in
the vicinity of the ferronematic $SU(3)$ point) for low-energy
processes the momenta of all the magnons participating in the process
are small, and one can use the long-wavelength asymptotic expressions
for the corresponding amplitudes:
\begin{widetext}
\begin{eqnarray}
\label{amplitudes}
\Phi_{1234} &=&
\frac{J_{2}^{3}}{32\sqrt{\varepsilon_{1}\varepsilon_{2}\varepsilon_{3}\varepsilon_{4}}}
\Big\{
 -k_{0}^{2}( 4k_{1} k_{2} k_{3} k_{4} +\mu_{1,2}^{+}\mu_{3,4}^{+})
+(\vec{k}_{1}\vec{k}_{3}+\vec{k}_{2}\vec{k}_{4})(\eta_{1,2}^{-}\eta_{3,4}^{-}-\mu_{1,2}^{-}\mu_{3,4}^{-})\nonumber\\
&& +(\vec{k}_{1}\vec{k}_{4}+\vec{k}_{2}\vec{k}_{3})(\eta_{1,2}^{-}\eta_{3,4}^{-}+\mu_{1,2}^{-}\mu_{3,4}^{-})
\Big\}\nonumber\\
\Psi_{1234} &=&
\frac{J_{2}^{3}}{16\sqrt{\varepsilon_{1}\varepsilon_{2}\varepsilon_{3}\varepsilon_{4}}}
\Big\{
 k_{0}^{2} (\mu_{1,3}^{-}\mu_{2,4}^{-} -4k_{1} k_{2} k_{3} k_{4})
-(\vec{k}_{1}\vec{k}_{2}+\vec{k}_{3}\vec{k}_{4}) (\eta_{1,3}^{+}\eta_{2,4}^{+}+\mu_{1,3}^{+}\mu_{2,4}^{+})\nonumber\\
&+&(\vec{k}_{1}\vec{k}_{4}+\vec{k}_{2}\vec{k}_{3})(\eta_{1,3}^{+}\eta_{2,4}^{+}-\mu_{1,3}^{+}\mu_{2,4}^{+})
\Big\}\\
F_{1234}&=&\frac{J_{2}^{3}}{32\sqrt{\varepsilon_{1}\varepsilon_{2}\varepsilon_{3}\varepsilon_{4}}}
\Big\{
k_{0}^{2}( 4k_{1}k_{2}k_{3}k_{4}+k_{1}k_{2}\eta_{3,4}^{+}+k_{3}k_{4}\eta_{1,2}^{+}-\mu_{1,2}^{+}\mu_{3,4}^{+})
+\frac{k^{2}}{2}
(\eta_{1,2}^{-}\eta_{3,4}^{-}+\mu_{1,2}^{-}\mu_{3,4}^{-})\nonumber\\
&-&2(\vec{k}_{1}\vec{k}_{2}+\vec{k}_{3}\vec{k}_{4})\mu_{1,3}^{+}\mu_{2,4}^{+}
-2(\vec{k}_{1}\vec{k}_{3}+\vec{k}_{2}\vec{k}_{4})\mu_{1,2}^{-}\mu_{3,4}^{-}
\Big\}.
\nonumber
 \end{eqnarray}
\end{widetext}
Here we have used the shorthand notation
\begin{eqnarray}
\label{shorthand2}
&& \eta_{n,m}^{\pm}=k_{n}k_{m}\pm \delta_{n}\delta_{m},\quad
  \mu_{n,m}^{\pm}=k_{n}\delta_{m}\pm\delta_{n}k_{m} \nonumber\\
&& \delta_{n}=\sqrt{k_{0}^{2}+k_{n}^{2}}, \quad k^{2}=k_{1}^{2}+k_{2}^{2}+k_{3}^{2}+k_{4}^{2}.
\end{eqnarray}

The magnon damping $\Gamma$ is determined as the imaginary
part of the self-energy calculated in the second order of perturbation
theory.
The expression for the damping can
be written in the following form:
\begin{eqnarray}
\label{GkT}
\Gamma_{\vec{k},T} &=&\frac{\pi}{2N^2}
\sinh\left( \frac{\varepsilon_{\vec{k}}} {2T} \right)
\sum_{\vec{p},\vec{q}} (\Phi^2 +2\Psi^2 +4F^{2}) \nonumber\\
&\times&\frac{ \delta \left( \varepsilon_{\vec{k}} +\varepsilon_{\vec{p}}
-\varepsilon_{\vec{q}} -\varepsilon_{\vec{k}+\vec{p}-\vec{q}}\right)}
{\sinh\left(\frac{\varepsilon_{\vec{p}}}{2T} \right)
\sinh\left(\frac{\varepsilon_{\vec{q}}}{2T} \right)
\sinh\left(\frac{\varepsilon_{\vec{k} +\vec{p}-\vec{q}}}{2T} \right)
},
\end{eqnarray}
where $T$ is the temperature in energy units, and the arguments of the
amplitudes (\ref{amplitudes}) are taken as
$\vec{k}_{1}=\vec{k}$, $\vec{k_{2}}=\vec{p}$, $\vec{k}_{3}=\vec{q}$, $\vec{k_{4}}=\vec{k}+\vec{p}-\vec{q}$.
We will consider the case of low excitation energies and
  temperatures well below the critical temperature $T_{c}\sim J_{2}$,
\begin{equation}
\label{highT}
\varepsilon_{\vec{k}}\ll T \ll J_{2},
\end{equation}
then the main contribution to the damping comes from thermally excited
magnons with wave vectors $p$ much higher than the magnon wave vector
$k$. This justifies neglecting magnon decays in the interaction Hamiltonian:
although such processes are allowed by the
energy and momentum conservation, their phase volume is
proportional to $k^{2}$ and thus their
contribution is small.

We will also assume that we are not too far away from the $SU(3)$
point, so that
\begin{equation}
\label{SU3close}
T\gg J_{2}k_{0}^{2}=2Z(J_{2}-J_{1}).
\end{equation}
At the $SU(3)$ point ($k_{0}=0$, $\delta_{n}=k_{n}$) the amplitude
$\Phi$ vanishes, and in the immediate vicinity of the $SU(3)$ point determined
by (\ref{SU3close}) it is much smaller than the other two amplitudes,
so $\Phi$ can be neglected in the calculation of the damping.

Then, in the leading order in $(J_{2}-J_{1})$ and $k/p$, one can
drop $k_{0}$  in the
dispersions of thermal magnons,
and the amplitudes in (\ref{GkT}) can be replaced by
\begin{equation}
\label{ampl-sim}
\Psi_{1234} =2F_{1234}=-\frac{J_{2}^{3/2}}{4\sqrt{\varepsilon_{1}}} ( k_{1}+\delta _{1})
(\vec{k}_{1}\cdot\vec{k}_{2}+ \vec{k}_{3}\cdot\vec{k}_{4}),
\end{equation}
where we have dropped contributions proportional to
$k_{0}^{2}/k_{1}k_{2} \ll 1$.
Further, if $k\gg k_{0}$, then $k_{0}$ can be neglected in the
equation for the mass surface as well, and then the result for the
damping under the assumptions
(\ref{highT}), (\ref{SU3close}) can be easily obtained as follows:
\begin{equation}
\label{G-micro-1}
\Gamma_{k,T}=\frac{J_{2}}{4} C_{k,T} k^{2}\left( 2k^{2}+k_{0}^{2} +O(k_{0}^{4})\right) \quad
\text{at}\quad  k_{0}^{2}
\ll k^{2} ,
\end{equation}
where $C_{k,T}$ is a dimensionless factor that weakly (logarithmically) depends on the wave vector $k$:
\begin{equation}
\label{C-micro-1}
C_{k,T} =\frac{T^{2}}{8\pi^3 J_{2}^{2}} \Big\{
\Big(\ln\frac{T}{\varepsilon_{k}}  +\frac{5}{3}\Big)^{2}   -\frac{4}{9}-(1-\ln 2)^2 \Big\}.
\end{equation}
At $\delta = 0$ this result reduces to the one found
before for the $SU(3)$ symmetric case.\cite{BIK10} The appearance of
the logarithm above is similar to the well known result for
isotropic ferromagnets.\cite{KaganovTsukernik58}

Comparing
(\ref{G-micro-1}) to the phenomenological result (\ref{Freq-su2-dyn}),
one can see that it is possible to match those two expressions if one
sets the correspondence
\begin{eqnarray}
\label{match-1}
&& \widetilde{\lambda}_{m}\approx \widetilde{\lambda}_{d}  \mapsto
C_{k,T},\quad \lambda_{d}=O\left((J_{2}-J_{1})^{2} \right),
\nonumber\\
&& \widetilde{\lambda}_{m}-\widetilde{\lambda}_{d}=O(J_{2}-J_{1}).
\end{eqnarray}
We see that our conjecture (\ref{lambda-d}), which stems from the
assumption that $\lambda_{d}$ is an analytical
function of $(J_{2}-J_{1})$, is confirmed by the results of the
microscopic calculations.

\section{Summary}
\label{sec:summary}

To summarize, we
have studied the properties of elementary excitations in the so-called
spin nematic phase of spin-$1$ systems that is
characterized by nontrivial quadrupole order in the absence of local spin averages.
We have developed a general phenomenological theory of spin dynamics and
relaxation for spin-$1$ systems, which is based on the equations of
motion for the eight-component real vector $\vec{n}$  uniting the magnetic (dipolar) and nematic (quadrupolar)
order parameters. Our approach is well suited to emphasize the role of enhanced
symmetry in the relaxation and is  in spirit similar to the
phenomenology describing relaxation in  ferromagnets.\cite{Baryakhtar84}

The developed theory has been applied to the specific $S=1$ lattice
model with isotropic bilinear and biquadratic exchange interactions,
which is relevant for the physics of ultracold spin-$1$ Bose gases in
optical lattices.  In the space of the model parameters, there is a
special point that exhibits an enhanced symmetry, namely, the usual
$SU(2)$ (rotational) symmetry is enhanced to $SU(3)$, and the
immediate vicinity of the $SU(3)$ point is occupied by the spin
nematic phase.  We show that the behavior of the leading term in the
dependence of the magnon damping on wavevector $k$ changes from
$k^{4}$ to $k^{2}$ as one moves away from the $SU(3)$ point and the
symmetry is lowered to $SU(2)$.
Those predictions of the phenomenological theory are consistent with the  results
of  microscopic calculations. We also show that breaking the $SU(3)$
symmetry leads to the appearance of a homogeneous ($k\to 0$)
relaxation in the  diffusive (non-propagating) quadrupolar mode that describes fluctuations
of $\langle S_{z}^{2}\rangle$.

\begin{acknowledgments}

This work has been partly supported by the  State Program
``Nanotechnologies and Nanomaterials'' of the Government of Ukraine,
Project 1.1.3.27, as well as by the joint Ukrainian-
 Russian research program via Grant No. 0113U001823 from
 the National Academy of Sciences and Grant No. F53.2/045
 from the State Foundation for Fundamental Research of
 Ukraine.

\end{acknowledgments}


\appendix*

\section{$SU(3)$ algebra and properties of octet vector products}
\label{app:su3}

For the sake of the reader's convenience, we list here the basic
algebraic relations \cite{Gell-Mann-8,Macfarlane+68} between the Gell-Mann matrices
$\lambda_{\alpha}$ and the structure tensors $f_{\alpha\beta\gamma}$,
$d_{\alpha\beta\gamma}$.

The  generators of the $SU(3)$ group can be chosen in the explicit matrix
representation  known as  Gell-Mann's matrices:
\begin{eqnarray}
\label{lambda-Gell-Mann}
&&\lambda_{1}=\begin{pmatrix} 0 & 1 & 0\\ 1 & 0 & 0\\ 0 & 0 &0 \end{pmatrix}, \
\lambda_{2}=\begin{pmatrix} 0 & -i & 0 \\ i & 0 & 0\\ 0 & 0 & 0 \end{pmatrix},\
\lambda_{3}=\begin{pmatrix} 1 & 0 & 0\\ 0 & -1 & 0 \\ 0 & 0 & 0 \end{pmatrix}, \nonumber\\
&& \lambda_{4}=\begin{pmatrix} 0 & 0 & 1 \\ 0 & 0 & 0 \\ 1 & 0 &0 \end{pmatrix}, \
\lambda_{5}=\begin{pmatrix} 0 & 0 & -i\\ 0 & 0 & 0\\ i & 0 &
0 \end{pmatrix}, \
\lambda_{6}=\begin{pmatrix} 0 & 0 & 0\\ 0 & 0 & 1 \\ 0 & 1 & 0 \end{pmatrix}, \nonumber\\
&& \lambda_{7}=\begin{pmatrix} 0 & 0 & 0\\ 0 & 0 & -i \\ 0 & i &
0 \end{pmatrix},\
\lambda_{8}=\frac{1}{\sqrt{3}}\begin{pmatrix} 1 & 0 & 0\\ 0 & 1 & 0\\ 0 & 0 & -2\end{pmatrix}
\end{eqnarray}
The algebraic properties of Gell-Mann's matrices are given by
Eq.\ (\ref{Gell-Mann}), with the totally antisymmetric structure tensor
$f_{\alpha\beta\gamma}$ and the totally symmetric structure tensor
$d_{\alpha\beta\gamma}$ defined by their  nonzero components as follows:
\begin{eqnarray}
\label{struct-f}
&& f_{123}=1,\quad f_{458}=f_{678}=\frac{\sqrt{3}}{2},\\
&&
f_{147}=f_{246}=f_{257}=f_{345}=f_{516}=f_{637}=\frac{1}{2},\nonumber
\end{eqnarray}
\begin{eqnarray}
\label{struct-d}
 d_{118}&=&d_{228}=d_{338}=-d_{888}=-\frac{1}{\sqrt{3}},\nonumber\\
 d_{448}&=&d_{558}=d_{668}=d_{778}=-\frac{1}{2\sqrt{3}},\\
 d_{146}&=&d_{157}=d_{256}=d_{344}=d_{355} \nonumber\\
&=&d_{247}=d_{366}=d_{377}=-\frac{1}{2}. \nonumber
\end{eqnarray}

With the help of the above structure constants, one can define the
symmetric and antisymmetric vector products as in
Eq.\ (\ref{octet-prod}).

If one splits the full space of octet vectors
into the magnetization subspace $V_{\vec{m}}$
and the quadrupolar subspace $V_{\vec{d}}$, see Eq.\ (\ref{md}),
then it is straightforward to check that the vector products have the properties
\begin{eqnarray}
\label{vecprods}
&&(\vec{d}\wedge \vec{d}') \in V_{\vec{m}},\ (\vec{d}\wedge \vec{m})
\in V_{\vec{d}},\
(\vec{m}\wedge \vec{m}') \in V_{\vec{m}}, \nonumber\\
&& (\vec{d} * \vec{d}') \in V_{\vec{d}},\ (\vec{d} * \vec{m}) \in
V_{\vec{m}},\
(\vec{m} * \vec{m}') \in V_{\vec{d}},
\end{eqnarray}
where $\vec{m},\vec{m}'\in V_{\vec{m}}$ and $\vec{d},\vec{d}'\in V_{\vec{d}}$.



\end{document}